\pdfoutput=1
\documentclass[a4paper,11pt]{article}
\usepackage[
top=30truemm,
bottom=30truemm,
left=25truemm,
right=25truemm
]{geometry} 
\usepackage[dvips]{graphicx}
\usepackage{amsmath,amsthm,amssymb}
\usepackage{amsfonts}
\usepackage{color}
\usepackage{pifont}
\usepackage{here}
\usepackage{cases}
\usepackage{empheq}
\usepackage{multirow} 
\usepackage{colortbl} 
\usepackage{arydshln} 

\def\Maximize{\displaystyle\mathop{\rm Maximize}}

\def\bs{\boldsymbol}
\def\dsum{\displaystyle\sum}
\def\cg{\cellcolor[gray]{0.90}} 
\begin{document}
\title{Optimal class assignment problem: \\
a case study at Gunma University}
\author{Akifumi~KIRA $^{\tt a, b, *}$, Kiyohito NAGANO $^{\tt a}$, Manabu SUGIYAMA $^{\tt a}$ \\
and Naoyuki KAMIYAMA $^{\tt b, c}$ \\
}
\date{\empty}
\maketitle
\vspace{-10mm}
\begin{center}
\small
\begin{tabular}{l}
$^{\tt a}$ Faculty of Social and Information Studies, Gunma University \\
\quad 4-2 Aramaki-machi, Maebashi, Gunma 371-8510, Japan \\
$^{\tt b}$ Institute of Mathematics for Industry, Kyushu University \\
\quad 744 Motooka, Nishi-ku, Fukuoka 819-0395, Japan \\
$^{\tt c}$ JST, PRESTO \\
\quad 4-1-8 Honcho, Kawaguchi, Saitama, 332-0012, Japan \\
$^{\tt *}$ Corresponding author, E-mail: a-kira@oak.gunma-u.ac.jp
\end{tabular}
\end{center}
\begin{abstract}
In this study, we consider the real-world problem of assigning students to classes, 
where each student has a preference list, ranking a subset of classes in order of preference. 
Though we use existing approaches to include the daily class assignment of Gunma University, 
new concepts and adjustments are required to find improved results depending on real instances 
in the field.   
Thus, we propose minimax-rank constrained maximum-utility matchings and 
a compromise between maximum-utility matchings and fair matchings, where a matching is 
said to be fair if it lexicographically minimizes the number of students 
assigned to classes not included in their choices, 
the number of students assigned to their last choices, and so on. 
In addition, we also observe the potential inefficiency of the student proposing deferred acceptance 
mechanism with single tie-breaking, which a hot topic in the literature on the school choice problem. 
\end{abstract}

\section{Introduction}
We can design 
fair and highly convincing systems and measures for social issues, 
by using mathematical optimization and game theory,
So far, the authors have also developed technology 
in collaboration with real world sites of social issues such as 
(a) efforts to improve passenger satisfaction at Fukuoka Airport~\cite{ohori2018mathematical, yamada2017modeling}, 
(b) system design for Japanese nursery school matching with siblings~\cite{kira2018nursery}, and 
(c) disaster restoration scheduling for lifeline networks~\cite{kira2017indirect}. 
The practice of operations research makes it possible to contribute to society. 
In particular, it enables us to contribute to the daily work of universities.

\par
In all schools and colleges, students are offered a plethora of subjects. 
Depending on their preferences and availability of seats, they are assigned to various classes. 
This is the classic assignment problem. 
At Gunma University, all first-year students (approximately 1100) must take a subject called 
\lq\lq Academic Literacy II" in the second semester. 
This subject is a compulsory elective subject in liberal arts education 
and consists of about 50 classes each year; each student must take one of these classes.   
Since the number of students in each class is limited, 
a questionnaire survey is conducted in advance asking students which classes they would like to take. 
Each student submits a preference list, ranking a subset of classes in order of preference. 
The university then assigns students to classes based on the preference lists so that 
the students can be satisfied.  
Until FY2017, the university outsourced this class assignment procedure to 
a system development company every year.
Approximately eighty percent of the students were assigned to classes up to their third choices; however, 
the remaining twenty percent of students were assigned to lower preferred classes (fourth to sixth choices), 
and many of the students were unsatisfied with the results. 
Thus, the authors were asked to improve this system.
\par
This type of matching problem, known as 
the class assignment problem, was introduced by Konno and Zhu~\cite{konno1991optimal}  
as a successful application of optimization theory. 
Their case study at the Tokyo Institute of Technology deals with 
the problem of assigning approximately 1200 students 
to 12 to 15 classes.  
They set the maximization of the sum of students' utilities as a criterion, 
and formulated the problem as a linear programming problem. 
It corresponds to the maximum-weight matching problem for a weighted bipartite graph. 
Konno~\cite{konno1992introduction} improved the model annually through discussions with students. 
\par 
In this study, we apply the optimization approach used by Konno to our class assignment problem. 
However, new concepts and adjustments are required to find improved results depending on real instances 
in the field.  
Hence, we propose minimax-rank constrained maximum-utility matchings to forcibly prohibit assignment to less preferred classes. 
We also propose a compromise between maximum-utility matching and fair matching. 
Specifically, a fair matching, first studied by Huang et al.~\cite{huang2016fair} as a 
derivative of rank-maximal matching~\cite{huang2019exact, irving2006rank, kavitha2006efficient, michail2007reducing, paluch2013capacitated}, 
is a matching that first minimizes the number of students 
assigned to classes not included in their choices. Subject to that constraint, a fair matching 
minimizes the number of students assigned to their last choices, and so on. 
We apply these methods to real instance and compare them with other well-known matching mechanisms.
From these experiments, we also observe the potential inefficiency of the student proposing deferred acceptance 
mechanism with single tie-breaking, which has been the hottest topic on the school choice problem 
since Erdil and Ergin~\cite{erdil2008what} pointed it out.  
In this study, the suitability of the optimization approach for class assignments 
will be confirmed.
%
\section{Basic Model} 
\label{Sec:case}
%
\lq\lq Academic Literacy II" is a compulsory elective subject in liberal arts education at Gunma University.
This subject consists of about 50 classes each year (see Table~\ref{Tbl:classes}), and each first-year student must take one class in the second semester.  
\begin{table}[H]
\centering
\caption{Classes and capacities}
\label{Tbl:classes}
\scalebox{1.0}{
\begin{tabular}{|c|l|c|c|} \hline
\multirow{2}{*}{Class} & \multirow{2}{*}{Title} & \multicolumn{2}{c|}{Capacity}  \\ \cline{3-4}
 &  & Min & Max   \\ \hline
$c_1$ & Reading Soseki Natsume & 7 & 40 \\
$c_2$ & Introduction to Physics & 7 & 25 \\
$c_3$ & Invitation to Modern Mathematics & 7 & 25 \\
$c_4$ & Let's Think about Environmental Issues. & 7 & 30 \\
$c_5$ & Learn ICT/IoT & 7 & 25 \\
$\vdots$  &  \qquad $\vdots$  &  $\vdots$  &  $\vdots$ \\
$c_{54}$ & History and Culture & 7 & 25 \\ \hline
\end{tabular}
}
\end{table}
 
\begin{figure}[H]
\centering
\begin{minipage}[b]{0.13\hsize}
\centering
$1$st \\
\includegraphics[width = 0.8\hsize]{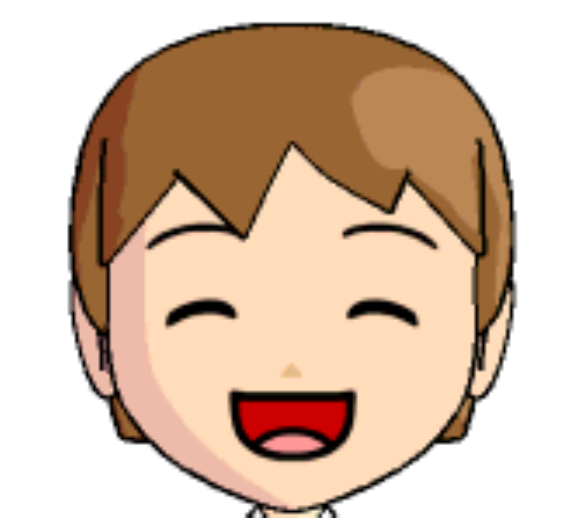} \\
$100$
\end{minipage}
\begin{minipage}[b]{0.13\hsize}
\centering
$2$nd \\
\includegraphics[width = 0.8\hsize]{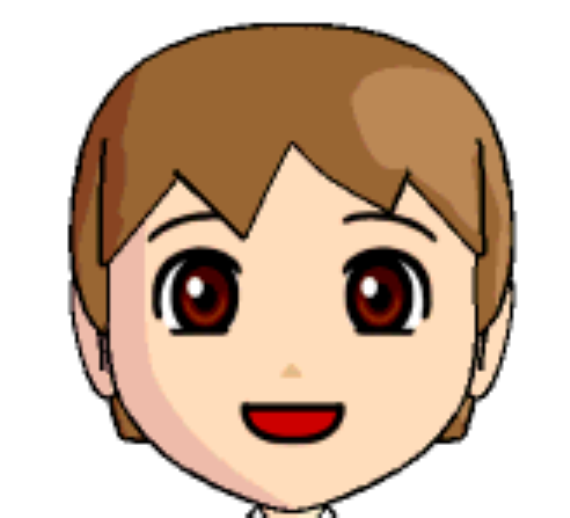} \\
$67$
\end{minipage}
\begin{minipage}[b]{0.13\hsize}
\centering
$3$rd \\
\includegraphics[width = 0.8\hsize]{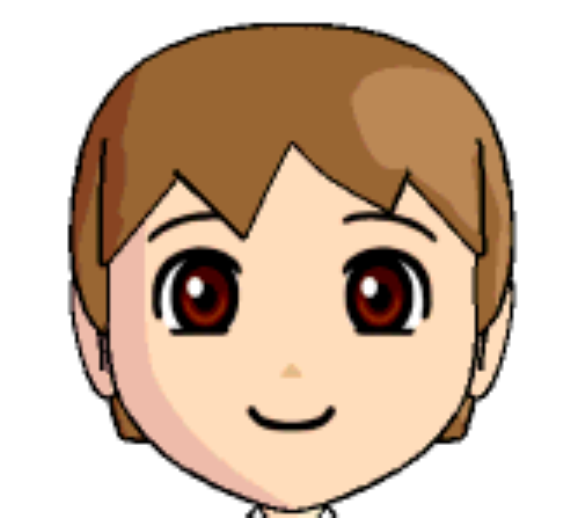} \\
$45$
\end{minipage}
\begin{minipage}[b]{0.13\hsize}
\centering
$4$th \\
\includegraphics[width = 0.8\hsize]{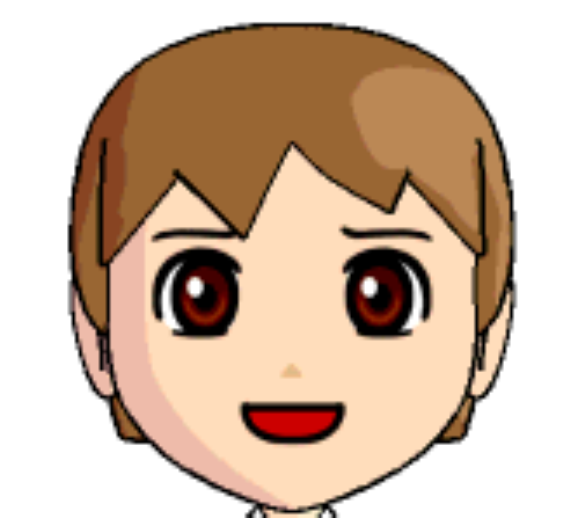} \\
$30$
\end{minipage}
\begin{minipage}[b]{0.13\hsize}
\centering
$5$th \\
\includegraphics[width = 0.8\hsize]{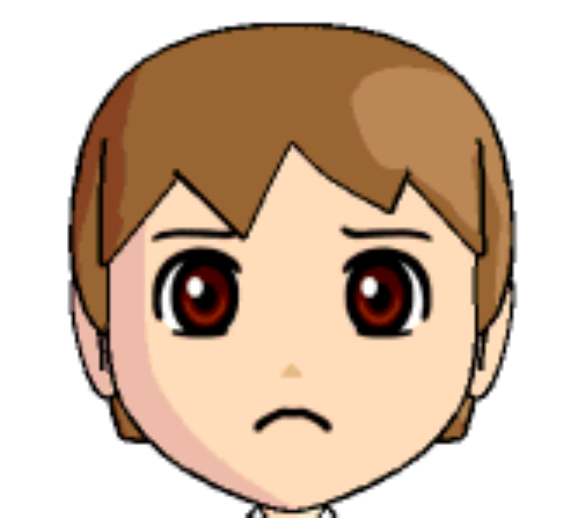} \\
$20$
\end{minipage}
\begin{minipage}[b]{0.13\hsize}
\centering
$6$th \\
\includegraphics[width = 0.8\hsize]{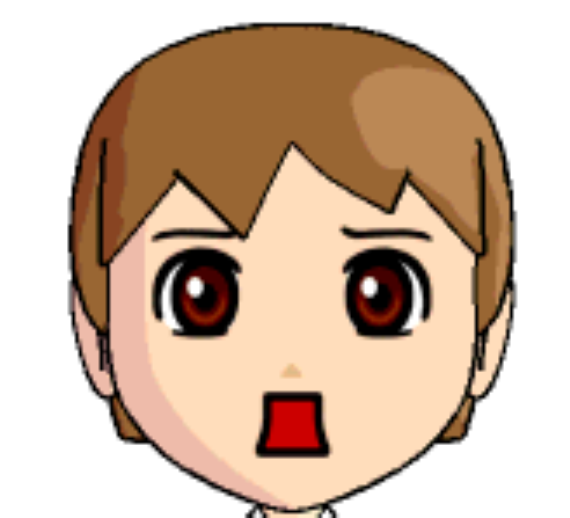} \\
$0$
\end{minipage}
\begin{minipage}[b]{0.13\hsize}
\centering
others \\
\includegraphics[width = 0.8\hsize]{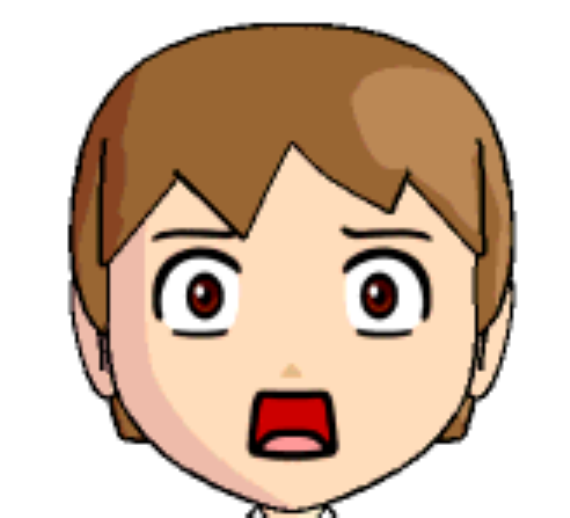} \\
$-10^6$
\end{minipage}
\caption{Utility of each student}
\label{fig:utility}
\end{figure}

For example, we assume the utility of each student as shown in Figure~\ref{fig:utility}.

Our goal is to find a matching that maximizes the sum of utilities (see Table~\ref{Tbl:matching}).

\begin{table}[H]
\centering
\caption{Total utility for a feasible matching}
\label{Tbl:matching}
\scalebox{1.0}{
\begin{tabular}{|c|c|c|c|c|c|c|c|c|} \hline
\multirow{2}{*}{Student} & \multicolumn{6}{c|}{Preference list}  & \multirow{2}{*}{Assigned} & \multirow{2}{*}{Utility} \\ \cline{2-7}
   & 1st & 2nd & 3rd & 4th & 5th & 6th &  &  \\ \hline
$s_1$ &  \cg $c_{30}$  &  $c_{25}$  &  $c_{51}$  &  $c_{42}$  &  $c_{32}$  &  $c_{35}$  &  $c_{30}$  &  100 \\
$s_2$ &  $c_{39}$  &  $c_{15}$  &  \cg $c_{16}$  &  $c_{02}$  &  $c_{01}$  &  $c_{33}$  &  $c_{16}$  &  45 \\
$s_3$ &  \cg $c_{33}$  &  $c_{16}$  &  $c_{05}$  &  $c_{37}$  &  $c_{03}$  &  $c_{34}$  &  $c_{33}$  &  100 \\
$\vdots$  &  $\vdots$  &  $\vdots$  &  $\vdots$  &  $\vdots$  &  $\vdots$  &  $\vdots$  &  $\vdots$  &  $\vdots$ \\
$s_{1138}$ &  $c_{51}$  &  \cg $c_{33}$  &  $c_{31}$  &  $c_{29}$  &  $c_{17}$  &  $c_{19}$  &  $c_{33}$  &  67 \\ \hline 
\multicolumn{8}{r}{Total utility} & \multicolumn{1}{c}{97684}  \\ 
\end{tabular}
}
\end{table}

Let $\mathcal{S}$ be the set of all students, and let $\mathcal{C}$ be the set of all classes. 
Each class $c$ has a capacity $u_c$, which refers to the maximum number of students that this class can accept. 
Furthermore, to prevent depopulated classes, 
all classes must accommodate a minimum of seven students.
Mathematically, a matching is a function $\mu : \mathcal{S} \to \mathcal{C}$ that satisfies 
\begin{gather*}
7 \leqq |\mu^{-1}(c)| \leqq u_c,
\end{gather*}
for all $c \in \mathcal{C}$, where,
\begin{equation*}
\mu^{-1}(c) := \{ s \in \mathcal{S} \,|\, \mu(s) = c \}.
\end{equation*}
We generalize the utility vector $(100, 67, 45, 30, 20, 0, -10^6)$ in Figure~\ref{fig:utility} as follows:
\begin{equation*}
\bs{p} = (p_1, p_2, \ldots, p_{\mathrm{others}}) \in \mathbb{R}^7.
\end{equation*}
For each $s \in \mathcal{S}$ and each $c \in \mathcal{C}$, 
we define an optimization variable $x_{sc}$ such that
\begin{equation*}
x_{sc} = 
\left\{
\begin{array}{cl}
1 & \textrm{if student $s$ is assigned to class $c$} \\
0 & \textrm{otherwise},
\end{array}
\right.
\end{equation*}
and define a constant $p_{sc}$ in the following manner:
\begin{equation*}
p_{sc} = 
\left\{
\begin{array}{cl}
p_1 & \textrm{if class $c$ is the $1$st choice of student $s$} \\
p_2 & \textrm{if class $c$ is the $2$nd choice of student $s$} \\
p_3 & \textrm{if class $c$ is the $3$rd choice of student $s$} \\
p_4 & \textrm{if class $c$ is the $4$th choice of student $s$} \\
p_5 & \textrm{if class $c$ is the $5$th choice of student $s$} \\
p_6 & \textrm{if class $c$ is the $6$th choice of student $s$} \\
p_{\mathrm{others}} & \textrm{otherwise}. 
\end{array}
\right.
\end{equation*}
Our problem can be expressed by the following integer programming problem. 
\begin{subequations}
\begin{empheq}[left = \mathrm{CA}(\boldsymbol{p}) \quad \empheqlvert \;\,]{align}
\Maximize \, \quad & \dsum_{s \in \mathcal{S}}  \dsum_{c \in \mathcal{C}} p_{sc} x_{sc}   \\[0mm]
\textrm{subject to} \quad  & \dsum_{c \in \mathcal{C}} x_{sc} = 1, \quad \forall s \in \mathcal{S} \\[0mm]
& \dsum_{s \in \mathcal{S}} x_{sc} \geqq 7, \quad \forall c \in \mathcal{C}  \label{lower-bound} \\[0mm]
& \dsum_{s \in \mathcal{S}} x_{sc} \leqq u_c, \quad \forall c \in \mathcal{C}  \\[0mm]
& x_{sc} \in \{0, 1\}, \quad \forall s \in \mathcal{S},\;\, \forall c \in \mathcal{C}. \label{binary}
\end{empheq}
\end{subequations}
Because the coefficients of the constraints form a totally unimodular matrix~\cite{schrijver1998theory},
we can replace the constraints (\ref{binary}) with
\begin{equation*}
0 \leqq x_{sc} \leqq 1, \quad \forall s \in \mathcal{S},\;\, \forall c \in \mathcal{C},
\end{equation*}
without loss of the objective value (e.g., see Conforti et al.~\cite[Theorem~4.5]{conforti2014integer}). 
We note that this linear programming (LP) problem is the so-called 
Hitchcock transportation problem.  
\par
The class assignment problem is closely related to the school choice problem
introduced by Abdulkadiro\u{g}lu and S\"{o}nmez~\cite{abdulkadiroglu2003school} 
as an application of matching theory. The important difference between the school choice problem 
and the class assignment problem is that in the school choice problem, each school 
has a priority ranking over students; 
in the class assignment problem, each class has no preference for the students. 
%
\section{Network flow formulation}
\begin{figure}[H]
\centering
\includegraphics[width = 120mm]{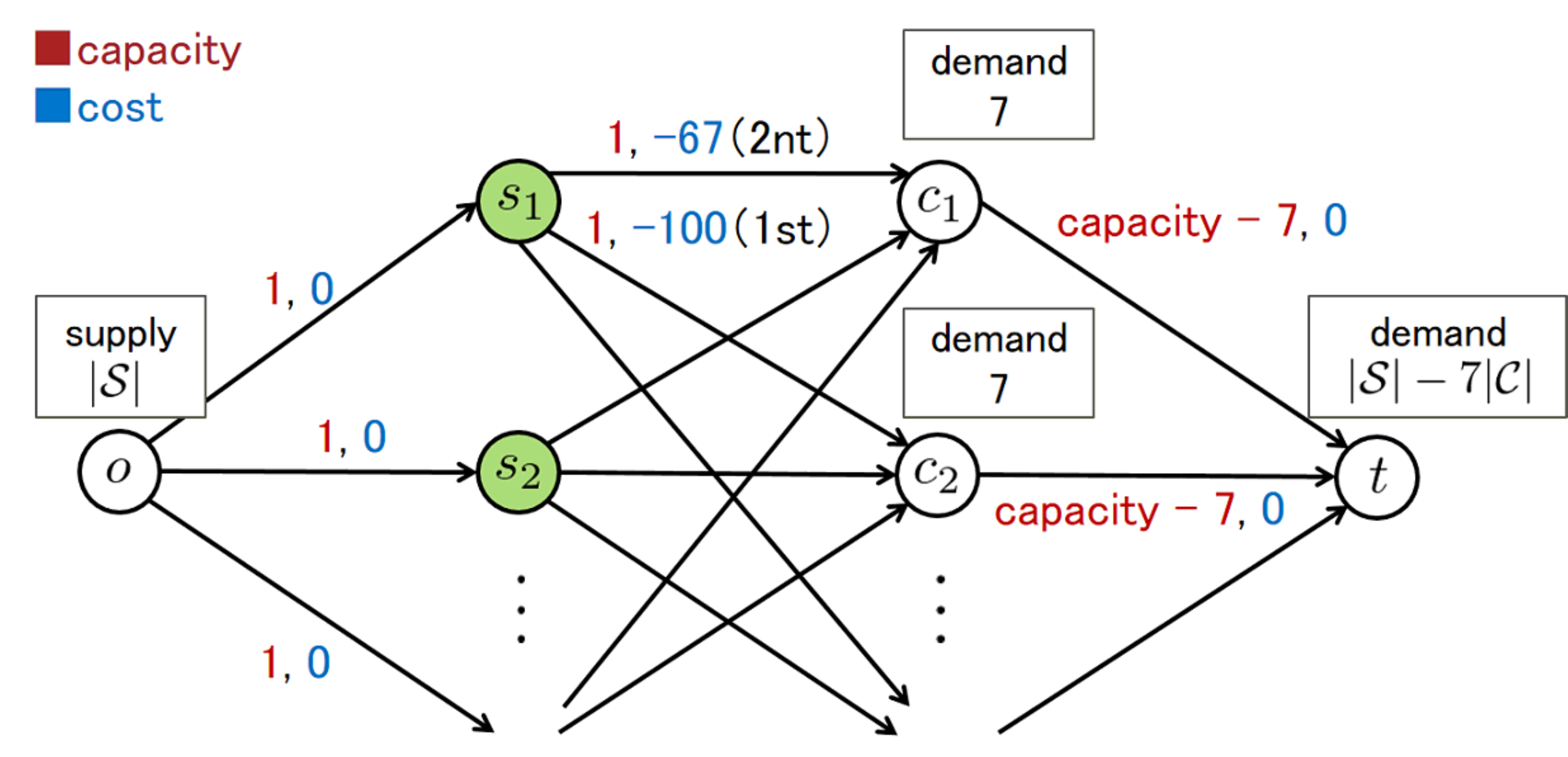}
\caption{Minimum-cost flow formulation of $\mathrm{CA}(\bs{p})$}
\label{Fig:mincost-flow}
\end{figure}

%
We can solve $\mathrm{CA}(\boldsymbol{p})$ by using a general purpose LP solver. 
Alternatively, since  $\mathrm{CA}(\boldsymbol{p})$ is a Hitchcock transportation problem, 
we can also convert it into a minimum-cost flow problem. 
We define the corresponding network $\mathcal{G} = (\mathcal{V}, \mathcal{E})$ with 
capacities $u : \mathcal{E} \to \mathbb{R}_{\geq 0}$, costs $h : \mathcal{E} \to \mathbb{R}$, and 
supplies/demands $b : \mathcal{V} \to \mathbb{R}$ as follows: 
For each student $s$ in $\mathcal{S}$, we create a node $s$. For each class $c$ in $\mathcal{C}$,  
we create a node $c$. 
Moreover, we add a super source $o$ and a super target $t$. Namely, 
\begin{equation*}
\mathcal{V} = \mathcal{S} \cup \mathcal{C} \cup \{ o, t \}.
\end{equation*}
The super source $o$ connects all student nodes and 
the super target $t$ is connected by each class node. 
For each $s$ in $\mathcal{S}$ and each $c$ in $\mathcal{C}$, there exists an arc that connects 
$s$ to $c$. Namely, 
\begin{equation*}
\mathcal{E} = (\{ o \} \times \mathcal{S}) \cup (\mathcal{S} \times \mathcal{C}) 
\cup (\mathcal{C} \cup  \{ t \}).
\end{equation*}
The capacity function $u$, the cost function $h$, and the supply/demand function $b$ are given by
\begin{align*}
u(e) &= \left\{
\begin{array}{cl}
u_{c} - 7 & \textrm{if $e = (c, t) \in \mathcal{C} \times \{ t \}$}, \\
1 & \textrm{otherwise}.
\end{array}
\right. \\
h(e) &= \left\{
\begin{array}{cl}
-p_{sc} & \textrm{if $e = (s, c) \in \mathcal{S} \times \mathcal{C}$}, \\
0 & \textrm{otherwise}.
\end{array}
\right. \\
b(v) &= \left\{
\begin{array}{cl}
|\mathcal{S}| & \textrm{if $v = o$},  \\
0 & \textrm{if $v \in \mathcal{S}$}, \\
-7 & \textrm{if $v \in \mathcal{C}$}, \\
7 |\mathcal{C}| - |\mathcal{S}|  & \textrm{if $v = t$}.
\end{array}
\right.
\end{align*}
Figure~\ref{Fig:mincost-flow} illustrates the network.

An $o$-$t$ $b$-flow is a function $f : \mathcal{E} \to \mathbb{R}$ satisfying
\begin{align*}
\dsum_{e \in \delta^{+}v} \! f(e) - \!
\dsum_{e \in \delta^{-}v} \! f(e) = b(e), \quad \forall v \in \mathcal{V}, \\
0 \leqq f(e) \leqq u(e), \quad \forall e \in \mathcal{E},
\end{align*}
where 
$\delta^{+} v$ and $\delta^{-}v$ denote the set of arcs $e \in \mathcal{E}$ 
leaving node $v$ and entering node $v$, respectively. 
For each $o$-$t$ $b$-flow $f$, the cost is given as follows:
\begin{equation*}
\sum_{e \in \mathcal{E}} h(e) f(e).
\end{equation*}
The goal of the minimum-cost flow problem is to find the $o$-$t$ $b$-flow that minimizes this value. 
It is well known that this setting of the minimum-cost flow problem has an integer flow solution. 
Namely, we can obtain the solution $f^*$ such that $f^*(e) \in \{0, 1\}$ for every $e \in \mathcal{S} \times \mathcal{C}$.
%
\section{Minimax-rank constrained matching}
%
We assign 1138 students to 54 classes in FY2008. 
Each student can choose 
only six classes they want to take and rank them.  
However, two of the classes were canceled  
after all the students submitted their preferences due to unforeseen circumstances.   
First, we solve the problem $\mathrm{CA}(\bs{p})$ for the four different utility vectors shown in 
Table~\ref{Tbl:4models_H30}, where $M$ is a sufficiently large integer and we set 
\begin{equation*}
M := 100 \times |\mathcal{S}| + 1.
\end{equation*}
In Opt80,  $\bs{p}$ is set so that $\frac{p_{i+1}}{p_{i}} \approx \frac{4}{5}$ for $i = 1,2,3,4$. 
In Opt75,  $\bs{p}$ is set so that $\frac{p_{i+1}}{p_{i}} \approx \frac{3}{4}$. 
In Opt67,  $\bs{p}$ is set so that $\frac{p_{i+1}}{p_{i}} \approx \frac{2}{3}$. 
In Opt50,  $\bs{p}$ is set so that $\frac{p_{i+1}}{p_{i}} \approx \frac{1}{2}$.  

\begin{table}[H]
\centering
\caption{Four models with different $\bs{p}$}
\label{Tbl:4models_H30}
\scalebox{1.0}{
\begin{tabular}{|l|l|} \hline
Model &  Utility vector $\bs{p}$ \\ \hline
Opt80 & $(100, 80, 64, 51, 41, 0, -M)$ \\
Opt75 & $(100, 75, 56, 42, 32, 0, -M)$ \\
Opt67 & $(100, 67, 45, 30, 20, 0, -M)$ \\
Opt50 & $(100, 50, 25, 13,  6, 0, -M)$ \\ \hline
\end{tabular}
}
\end{table}

In the case study at the Tokyo Institute of technology, 
Konno~\cite{konno1992introduction} finally proposes the procedure for each student to declare her/his 
utility vector. Applying this method to our case, each student must submit 
$\bs{p}_s = (p_1, p_2, \ldots, p_{\textrm{others}})$ satisfying
\begin{equation*}
p_1 = 100, \quad p_i \in [0, 100], \;\, i = 2,3, \ldots,6, \quad p_{\textrm{others}} = -M.
\end{equation*}
However, when discussing this with staffs of the Liberal Arts Education Section, 
they were reluctant to follow this approach, 
They believed that students may submit their choices without a full understanding of the procedure despite it being explained in advance. 
Therefore, we determined the value of $\bs{p}$ without input from the students.
\par
\begin{figure}[H]
\centering
\includegraphics[width = 130mm]{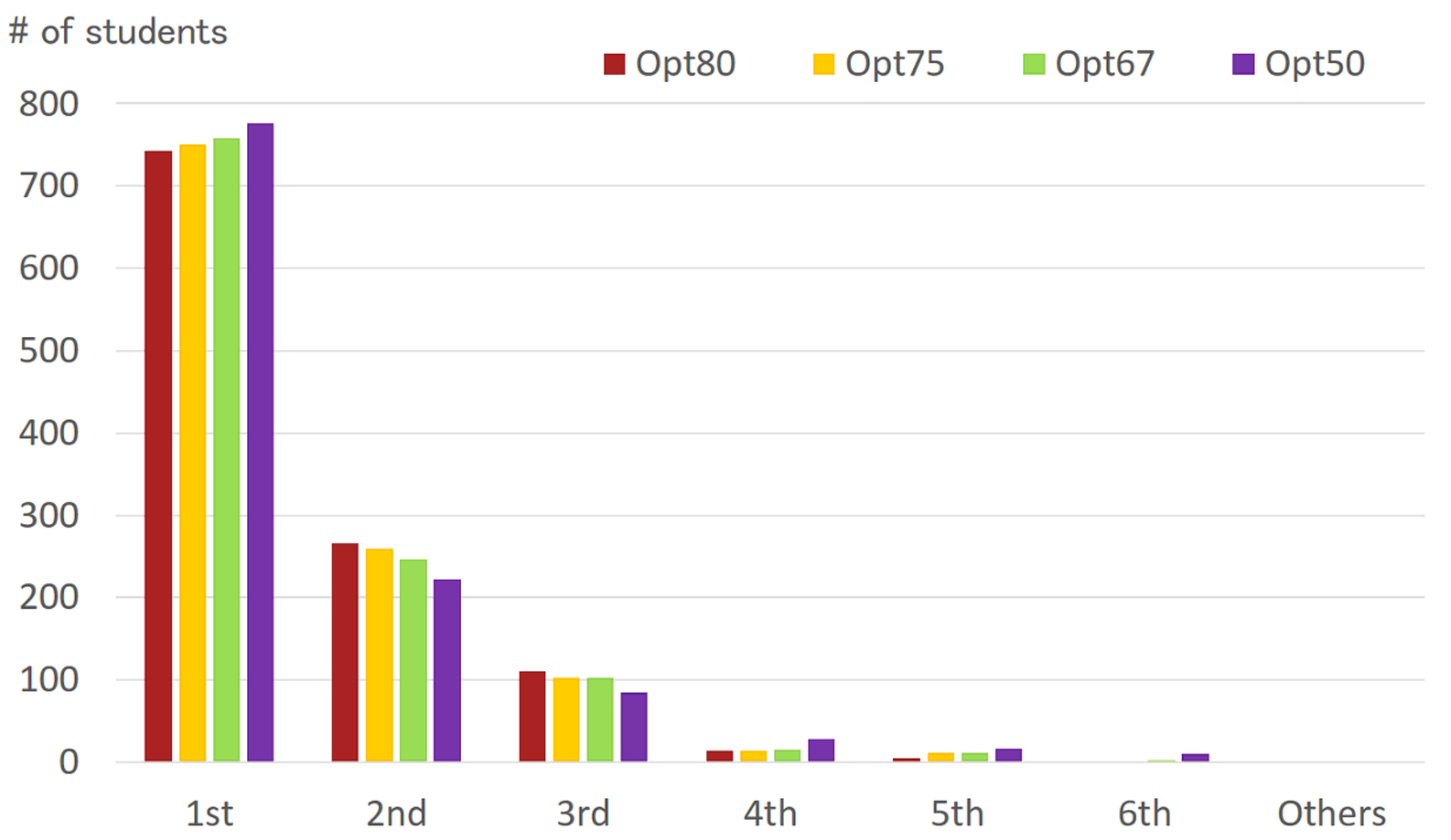}
\caption{Matching results by the four models with different $\bs{p}$ in FY2018}
\label{Fig:results1_H30}
\end{figure}

We implemented our solver using Python and NetworkX~\cite{hagberg2008exploring}, 
and executed it on a laptop PC equipped with an Intel\textregistered
      ~Core\texttrademark~i7-8565U processor and 16GB memory.
The computation of the minimum-cost flow for each $\bs{p}$ was completed in 
five seconds.  
The results are shown in Table~\ref{Tbl:results1_H30} and Figure~\ref{Fig:results1_H30}.

\begin{table}[H]
\centering
\caption{Matching results by the four models with different $\bs{p}$ in FY2018}
\label{Tbl:results1_H30}
\scalebox{1.0}{
\begin{tabular}{|l|c|c|c|c|c|c|c|c|c|} \hline
\multirow{2}{*}{Model}  &  \multicolumn{7}{c|}{\# of students (Total = 1138)}  &  \multicolumn{2}{c|}{Average}  \\ \cline{2-10}
  &  1st  &  2nd  &  3rd  &  4th  &  5th  &  6th  &  Others  &  Utility &  Rank \\ \hline
Opt80 &  743  &  266  &  110  &  14  &  5  &  0  &  0  &  90.983  &  1.481 \\
Opt75 &  750  &  259  &  103  &  14  &  12  &  0  &  0  &  88.897  &  1.487 \\
Opt67 &  758  &  247  &  103  &  16  &  11  &  3  &  0  &  85.838  &  1.492 \\
Opt50 &  776  &  222  &  85  &  28  &  17  &  10  &  0  &  80.220  &  1.521 \\ \hline
\end{tabular}
}
\end{table}

\begin{figure}[H]
\centering
\includegraphics[width = 130mm]{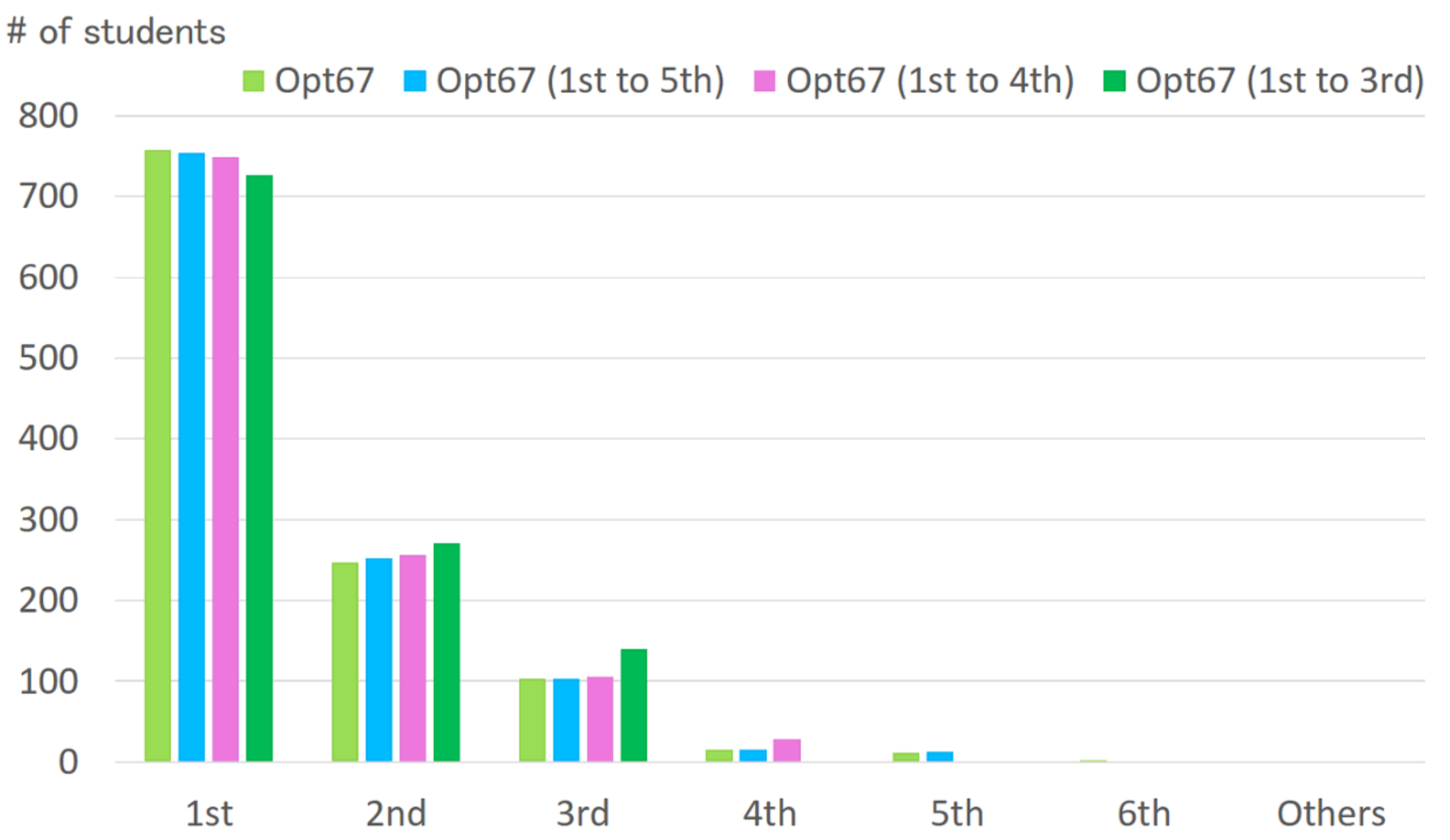}
\caption{Matching results by the restricted versions of Opt67 in FY2018}
\label{Fig:results2_H30}
\end{figure}

The results for Opt80 and Opt75 are better 
than that of the other models in that no students are assigned to their sixth choices.
However, in these cases, the difference in utility between the $i$th and $(i+1)$st choices is small.
Thus, Opt65 and Opt50 may be preferable in that we can reflect the wishes of students who strongly desire to be assigned to higher preferred classes. Therefore, while adopting the utility vector of Opt65, 
we attempt to forcibly prohibit assignment to lower preferred classes.  
We can easily accomplish this by replacing the utilities of the classes with $-M$, 
as in Table~\ref{Tbl:restricted_H30}. 
The results are shown in Table~\ref{Tbl:results2_H30}, and Figure~\ref{Fig:results2_H30}. 
The final assignments are shown in Table~\ref{Tbl:final_H30}. 

\begin{table}[H]
\centering
\caption{Restricted versions of Opt67}
\label{Tbl:restricted_H30}
\scalebox{1.0}{
\begin{tabular}{|l|l|} \hline
Model &  Utility vector $\bs{p}$ \\ \hline
Opt67 (1st to 5th) & $(100, 67, 45, 30, 20, -M, -M)$ \\
Opt67 (1st to 4th) & $(100, 67, 45, 30, -M, -M, -M)$ \\
Opt67 (1st to 3rd) & $(100, 67, 45, -M, -M, -M, -M)$ \\
Opt67 (1st to 2nd) & $(100, 67, -M, -M, -M, -M, -M)$ \\ \hline
\end{tabular}
}
\end{table}

\begin{table}[H]
\centering
\caption{Matching results by the restricted versions of Opt67 in FY2018}
\label{Tbl:results2_H30}
\scalebox{1.0}{
\begin{tabular}{|l|c|c|c|c|c|c|c|c|c|} \hline
\multirow{2}{*}{Model}  &  \multicolumn{7}{c|}{\# of students (Total = 1138)}  &  \multicolumn{2}{c|}{Average}  \\ \cline{2-10}
  &  1st  &  2nd  &  3rd  &  4th  &  5th  &  6th  &  Others  &  Utility &  Rank \\ \hline
Opt67 &  758  &  247  &  103  &  16  &  11  &  3  &  0  &  85.838  &  1.492 \\
Opt67 (1st to 5th) &  754  &  252  &  103  &  16  &  13  &  0  &  0  &  85.816  &  1.49 \\
Opt67 (1st to 4th) &  748  &  256  &  106  &  28  &  0  &  0  &  0  &  85.731  &  1.485 \\
Opt67 (1st to 3rd)  &  727  &  271  &  140  &  0  &  0  &  0  &  0  &  85.375  &  1.484 \\ \hdashline
Opt67 (1st to 2nd)  &  \multicolumn{9}{c|}{infeasible}  \\ \hline
\end{tabular}
}
\end{table}

\begin{table}[p]
\centering
\caption{Final assignment by Opt67 (1st to 3rd) in FY2018}
\label{Tbl:final_H30}
\scalebox{0.82}{
\begin{tabular}{|r|r|r|r|r|r|r|r|r|r|r|r|r|r|r|} \hline
\multirow{2}{*}{Class} & \multicolumn{2}{c|}{Capacity}  &  \multicolumn{5}{c|}{\cg \# of students assigned} &  \multicolumn{7}{c|}{\cg \# of students ranking the class} \\ \cline{2-3} \cline{5-8} \cline{10-15}
 & Lower & Upper & \cg Total & E & M & SS & ST & \cg Total & 1st & 2nd & 3rd & 4th & 5th & 6th \\ \hline
1 & 7 & 25 & \cg 25 & 11 & 8 & 2 & 4 & \cg 93 & 4 & 8 & 22 & 15 & 22 & 22 \\ 
2 & 7 & 40 & \cg 40 & 22 & 12 & 3 & 3 & \cg 198 & 94 & 22 & 13 & 20 & 21 & 28 \\ 
3 & 7 & 30 & \cg 30 & 6 & 5 & 6 & 13 & \cg 147 & 20 & 25 & 27 & 18 & 26 & 31 \\ 
4 & \multicolumn{2}{c|}{Canceled} & \cg 0 & 0 & 0 & 0 & 0 & \cg 69 & 17 & 10 & 10 & 14 & 9 & 9 \\ 
5 & 7 & 25 & \cg 25 & 2 & 0 & 8 & 15 & \cg 90 & 20 & 17 & 14 & 19 & 5 & 15 \\ 
6 & 7 & 20 & \cg 7 & 0 & 2 & 1 & 4 & \cg 19 & 1 & 4 & 3 & 2 & 1 & 8 \\ 
7 & 7 & 16 & \cg 7 & 0 & 1 & 0 & 6 & \cg 56 & 3 & 7 & 6 & 11 & 11 & 18 \\ 
8 & 7 & 26 & \cg 26 & 1 & 6 & 1 & 18 & \cg 123 & 15 & 25 & 17 & 19 & 20 & 27 \\ 
9 & 7 & 25 & \cg 25 & 10 & 4 & 5 & 6 & \cg 164 & 14 & 33 & 28 & 35 & 31 & 23 \\ 
10 & 7 & 25 & \cg 25 & 4 & 0 & 5 & 16 & \cg 200 & 22 & 30 & 42 & 37 & 31 & 38 \\ 
11 & 7 & 25 & \cg 7 & 1 & 2 & 0 & 4 & \cg 39 & 2 & 8 & 8 & 9 & 8 & 4 \\ 
12 & 7 & 25 & \cg 25 & 9 & 0 & 2 & 14 & \cg 175 & 17 & 43 & 20 & 34 & 34 & 27 \\ 
13 & 7 & 25 & \cg 25 & 11 & 2 & 3 & 9 & \cg 147 & 20 & 30 & 17 & 22 & 32 & 26 \\ 
14 & 7 & 25 & \cg 25 & 10 & 0 & 0 & 15 & \cg 270 & 80 & 60 & 39 & 40 & 28 & 23 \\ 
15 & 7 & 25 & \cg 12 & 4 & 0 & 4 & 4 & \cg 44 & 4 & 7 & 9 & 6 & 10 & 8 \\ 
16 & 7 & 25 & \cg 24 & 0 & 7 & 0 & 17 & \cg 95 & 13 & 15 & 6 & 18 & 16 & 27 \\ 
17 & 7 & 30 & \cg 24 & 3 & 0 & 0 & 21 & \cg 88 & 16 & 9 & 16 & 19 & 10 & 18 \\ 
18 & 7 & 20 & \cg 20 & 3 & 4 & 9 & 4 & \cg 101 & 15 & 19 & 15 & 16 & 18 & 18 \\ 
19 & 7 & 25 & \cg 22 & 11 & 1 & 2 & 8 & \cg 138 & 12 & 7 & 31 & 41 & 23 & 24 \\ 
20 & 7 & 40 & \cg 40 & 1 & 18 & 2 & 19 & \cg 256 & 69 & 49 & 56 & 30 & 28 & 24 \\ 
21 & 7 & 25 & \cg 9 & 2 & 3 & 1 & 3 & \cg 57 & 0 & 7 & 8 & 6 & 18 & 18 \\ 
22 & 7 & 30 & \cg 30 & 8 & 3 & 5 & 14 & \cg 202 & 15 & 29 & 36 & 35 & 58 & 29 \\ 
23 & 7 & 30 & \cg 30 & 10 & 4 & 1 & 15 & \cg 127 & 18 & 20 & 21 & 25 & 24 & 19 \\ 
24 & 7 & 30 & \cg 30 & 20 & 3 & 2 & 5 & \cg 105 & 18 & 20 & 16 & 10 & 22 & 19 \\ 
25 & \multicolumn{2}{c|}{Canceled} & \cg 0 & 0 & 0 & 0 & 0 & \cg 176 & 41 & 34 & 37 & 18 & 24 & 22 \\ 
26 & 7 & 40 & \cg 40 & 0 & 25 & 0 & 15 & \cg 263 & 50 & 59 & 45 & 42 & 38 & 29 \\ 
27 & 7 & 25 & \cg 10 & 2 & 2 & 0 & 6 & \cg 78 & 2 & 7 & 10 & 14 & 21 & 24 \\ 
28 & 7 & 25 & \cg 25 & 1 & 5 & 0 & 19 & \cg 280 & 88 & 47 & 51 & 36 & 34 & 24 \\ 
29 & 7 & 40 & \cg 30 & 4 & 8 & 3 & 15 & \cg 118 & 4 & 19 & 19 & 22 & 22 & 32 \\ 
30 & 7 & 18 & \cg 16 & 5 & 3 & 6 & 2 & \cg 78 & 10 & 13 & 9 & 10 & 19 & 17 \\ 
31 & 7 & 25 & \cg 25 & 0 & 19 & 1 & 5 & \cg 269 & 36 & 51 & 46 & 41 & 45 & 50 \\ 
32 & 7 & 24 & \cg 24 & 0 & 8 & 1 & 15 & \cg 120 & 6 & 20 & 24 & 23 & 28 & 19 \\ 
33 & 7 & 25 & \cg 25 & 1 & 19 & 0 & 5 & \cg 182 & 43 & 22 & 34 & 33 & 17 & 33 \\ 
34 & 7 & 25 & \cg 25 & 4 & 14 & 1 & 6 & \cg 175 & 15 & 30 & 26 & 38 & 42 & 24 \\ 
35 & 7 & 30 & \cg 13 & 7 & 2 & 3 & 1 & \cg 55 & 5 & 6 & 14 & 9 & 7 & 14 \\ 
36 & 7 & 15 & \cg 14 & 8 & 2 & 2 & 2 & \cg 53 & 3 & 9 & 6 & 11 & 15 & 9 \\ 
37 & 7 & 25 & \cg 10 & 1 & 2 & 2 & 5 & \cg 56 & 2 & 7 & 11 & 15 & 10 & 11 \\ 
38 & 7 & 20 & \cg 12 & 1 & 7 & 0 & 4 & \cg 66 & 4 & 5 & 15 & 12 & 15 & 15 \\ 
39 & 7 & 35 & \cg 35 & 7 & 12 & 1 & 15 & \cg 216 & 28 & 32 & 34 & 48 & 34 & 40 \\ 
40 & 7 & 20 & \cg 7 & 2 & 3 & 0 & 2 & \cg 31 & 3 & 1 & 4 & 10 & 4 & 9 \\ 
41 & 7 & 25 & \cg 14 & 0 & 1 & 0 & 13 & \cg 50 & 8 & 10 & 7 & 8 & 8 & 9 \\ 
42 & 7 & 25 & \cg 25 & 12 & 4 & 6 & 3 & \cg 99 & 23 & 16 & 12 & 13 & 11 & 24 \\ 
43 & 7 & 20 & \cg 20 & 2 & 3 & 0 & 15 & \cg 169 & 28 & 18 & 40 & 29 & 25 & 29 \\ 
44 & 7 & 25 & \cg 24 & 1 & 0 & 3 & 20 & \cg 102 & 16 & 12 & 14 & 14 & 15 & 31 \\ 
45 & 7 & 40 & \cg 40 & 7 & 16 & 1 & 16 & \cg 167 & 43 & 32 & 29 & 26 & 20 & 17 \\ 
46 & 7 & 16 & \cg 16 & 3 & 5 & 0 & 8 & \cg 53 & 12 & 7 & 5 & 11 & 10 & 8 \\ 
47 & 7 & 20 & \cg 20 & 8 & 4 & 0 & 8 & \cg 107 & 26 & 17 & 19 & 23 & 8 & 14 \\ 
48 & 7 & 40 & \cg 29 & 1 & 3 & 4 & 21 & \cg 87 & 12 & 16 & 15 & 15 & 20 & 9 \\ 
49 & 7 & 30 & \cg 7 & 0 & 0 & 4 & 3 & \cg 52 & 3 & 6 & 11 & 9 & 9 & 14 \\ 
50 & 7 & 25 & \cg 25 & 0 & 8 & 0 & 17 & \cg 272 & 53 & 51 & 47 & 45 & 45 & 31 \\ 
51 & 7 & 20 & \cg 20 & 11 & 3 & 0 & 6 & \cg 212 & 32 & 48 & 33 & 31 & 32 & 36 \\ 
52 & 7 & 25 & \cg 25 & 0 & 1 & 1 & 23 & \cg 133 & 18 & 20 & 24 & 18 & 32 & 21 \\ 
53 & 7 & 20 & \cg 13 & 2 & 1 & 2 & 8 & \cg 52 & 7 & 11 & 10 & 5 & 8 & 11 \\ 
54 & 7 & 25 & \cg 16 & 2 & 2 & 0 & 12 & \cg 54 & 8 & 8 & 7 & 8 & 14 & 9 \\ 
\hline
\end{tabular}
} \\
{\footnotesize 
\begin{itemize}
\item E = Faculty of Education, M = Faculty of Medicine, SS = Faculty of Social and Information Studies, 
ST = School of Science and Technology
\item The order of the classes has been shuffled to ensure anonymity.
\end{itemize}
}
\end{table}

As a result, we succeed in assigning all students to classes up to their third choice. 
Compared with the conventional methods in which approximately 20 percent 
of students were assigned to lower preferred classes (fourth to sixth choices), 
this result shows a significant improvement.
The staffs of the Liberal Arts Education Section were deeply impressed by the efficacy 
of the optimization theory and decided to adopt Opt67 (1st to 3rd) as the final class assignment.  
%
\section{Optimization v.s. Boston and DA}
\begin{figure}[H]
\centering
\includegraphics[width = 130mm]{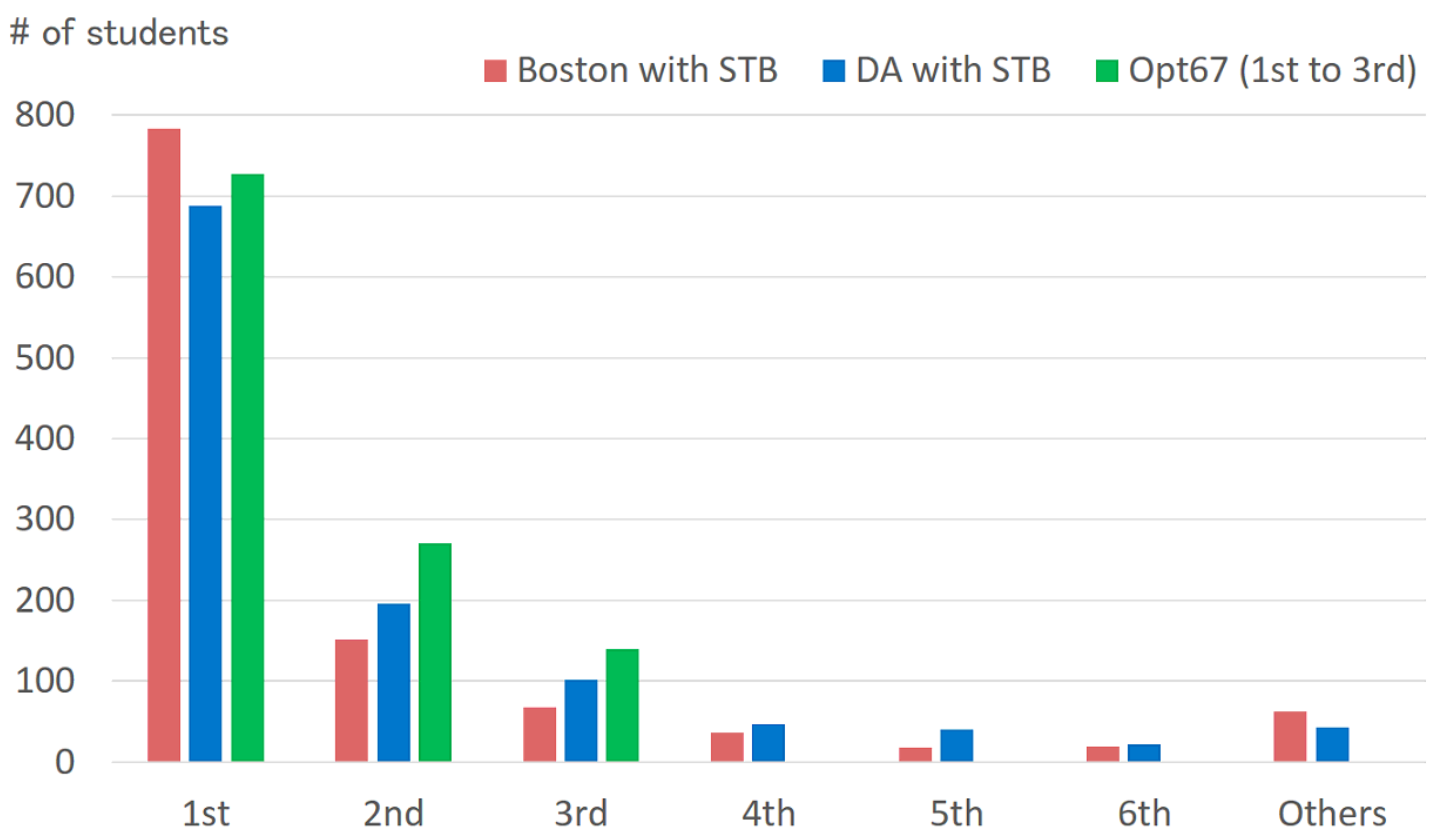}
\caption{Matching results by the DA and Boston mechanisms in FY2018}
\label{Fig:results3_H30}
\end{figure}

%
We succeeded in assigning all students to classes up to their third choices. 
Nevertheless, it was argued that the Gale--Shapley deferred acceptance (DA) algorithm~\cite{gale1962college}, which is a promising mechanism for the school choice problem, 
should be used instead of the optimization approach. 
If we ignore the lower capacity constraints~(\ref{lower-bound}), and 
if we use a single tie-breaking procedure (STB), that is, if we shuffle the student list 
and regard it as a priority ranking over students, then we obtain a situation in which 
the student-proposing DA can be applied.\footnote{For stable matchings with lower quotas, see e.g.,  
Bir\`{o} et al.~\cite{biro2010college}, Fleiner and Kamiyama~\cite{fleiner2016matroid}, 
Fragiadakis et al.~\cite{fragiadakis2016}, Hamada et al.~\cite{hamada2016hospitals}, 
Huang~\cite{huang2010classified}, and Yokoi~\cite{yokoi2017generalized}.}
Therefore, we conducted additional experiments. The results of applying 
the student proposing DA and 
the Boston mechanisms,\footnote{The Boston mechanism is a popular student-placement mechanism in school-choice 
programs around the world. In Faculty of Social and Information Studies, Gunma University, 
to which three of authors belong, the Boston mechanism is also used 
when students select seminars for their third year. } both with STB 
 to our problem are shown in Table~\ref{Tbl:results3_H30} and Figure~\ref{Fig:results3_H30}.

\begin{table}[H]
\centering
\caption{Matching results by the DA and Boston mechanisms in FY2018}
\label{Tbl:results3_H30}
\scalebox{1.0}{
\begin{tabular}{|l|c|c|c|c|c|c|c|c|c|} \hline
\multirow{2}{*}{Model}  &  \multicolumn{7}{c|}{\# of students (Total = 1138)}  &  \multicolumn{2}{c|}{Average}  \\ \cline{2-10}
  &  1st  &  2nd  &  3rd  &  4th  &  5th  &  6th  &  Others  &  Utility &  Rank \\ \hline
Boston with STB  &  783  &  151  &  68  &  36  &  18  &  19  &  63  &  --  &  $\geqq$ 1.860 \\
DA with STB  &  688  &  196  &  102  &  47  &  40  &  22  &  43  &  --  &  $\geqq$ 1.939 \\ \hline
\end{tabular}
}
\end{table}

\par 
Since the lower capacity constraints are not considered, 
five and four classes do not reach seven students in Boston with STB and 
DA with STB, respectively.  
Nevertheless, a considerable number of 
students are assigned to classes not included in their choices. 
Thus, these results could never be considered satisfactory. 
\par
Certainly, in the school choice problem, the student-proposing DA is strategy-proof for students, 
meaning that no student has an incentive to misstate her/his true preference list. 
Furthermore, it yields a student optimal stable matching within all stable matchings;
that is, it yields a stable matching that is not Pareto dominated by any other stable matching. 
By contrast, the optimization approach is not strategy-proof for students, 
but it yields a student optimal matching in all (not necessarily stable) matchings. 
\par
In fact, the inefficiency of the DA mechanism with random tie-breaking is known in the 
literature on the school choice problem. 
According to Erdil and Ergin~\cite{erdil2008what}, 
versions of the DA algorithm are being adopted in several school choice districts in
the U.S. To their knowledge, all of them employ a random tie-breaking rule when
faced with indifferences in the priority orders. 
Erdil and Ergin~\cite{erdil2008what} note that 
this may cause a significant loss of efficiency, because such a random tie-breaking rule introduces 
artificial stability constraints. 
Abdulkadiro\u{g}lu et al.~\cite{abdulkadiroglu2009strategy} prove that 
there exists no strategy-proof mechanism (stable or otherwise) which Pareto improves on the DA  
with single tie-breaking,
and state that the potential inefficiency of the DA with single tie-breaking is the cost of strategy-proofness. 
Furthermore, Abdulkadiro\u{g}lu et al.~\cite{abdulkadiroglu2011resolving} show a problem setting in which  
the Boston mechanism with random tie-breaking Pareto dominates the DA mechanism 
with random tie-breaking in ex ante welfare (i.e., in the sense of 
expected utility). This is interesting because it is in contrast to the fact 
shown by Ergin and S{\"o}nmez~\cite{ergin2006games}  
that the Boston mechanism is (weakly) Pareto dominated by the DA mechanism when priorities are strict.
%
\section{Maximum-utility matching and fair matching}
\begin{figure}[H]
\centering
\includegraphics[width = 120mm]{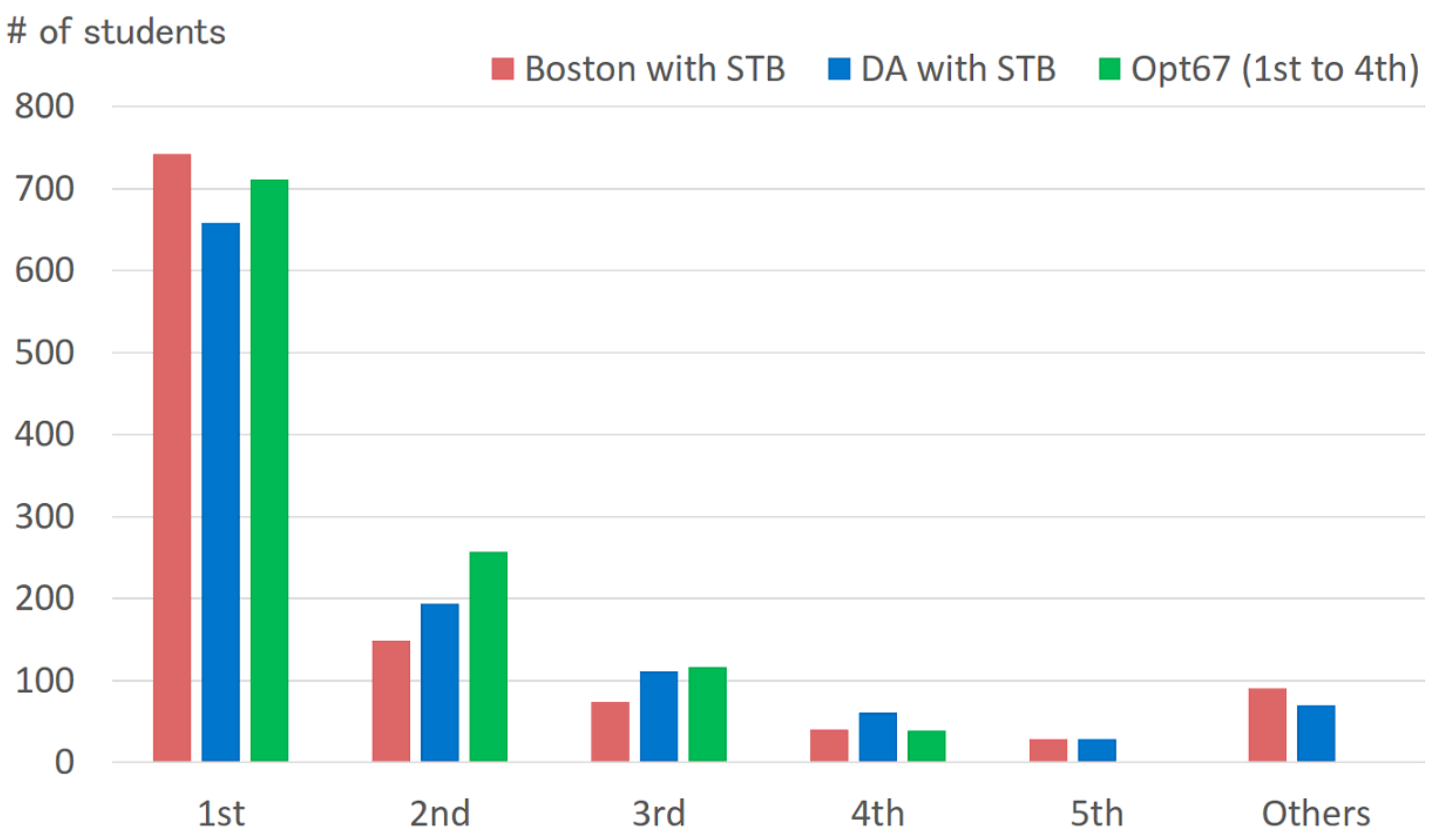}
\caption{Matching results by the restricted versions of Opt67 in FY2019}
\label{Fig:results1_R1}
\end{figure}

%
We assigned 1123 students to 50 classes in FY2019. 
Based on the successful results in FY2018, 
the staffs of the Liberal Arts Education Section decided to reduce the number of classes ranked by each student from six to five. 
As in FY2018, we solved the restricted versions of Opt67 
and compared the results with the matching results of the Boston mechanism with STB and 
the student proposing DA mechanism with STB. 
The results are shown in Table~\ref{Tbl:results1_R1} and Figure~\ref{Fig:results1_R1}.

\begin{table}[H]
\centering
\caption{Matching results by the restricted versions of Opt67 in FY2019}
\label{Tbl:results1_R1}
\scalebox{1.0}{
\begin{tabular}{|c|c|c|c|c|c|c|c|c|} \hline
\multirow{2}{*}{Model}  &  \multicolumn{6}{c|}{\# of students (Total = 1123)}  &  \multicolumn{2}{c|}{Average}  \\ \cline{2-9}
  &  1st  &  2nd  &  3rd  &  4th  &  5th  &  Others  &  Utility &  Rank \\ \hline
Boston with STB  &  742  &  148  &  74  &  40  &  29  &  90  &  --  &  $\geqq$ 1.874 \\
DA with STB  &  659  &  193  &  111  &  61  &  29  &  70  &  --  &  $\geqq$ 1.947 \\
Opt67 (1st to 4th)   &  711  &  257  &  116  &  39  &  0  &  0  &  84.34   &  1.540 \\ \hdashline
Opt67 (1st to 3rd)  &  \multicolumn{8}{c|}{infeasible}  \\ \hline
\end{tabular}
}
\end{table}

We note again that the lower capacity constraints were not considered both in Boston with STB and 
DA with STB. As a result,  
four classes and three classes did not reach seven students in each scenario, respectively.  
\par
Contrary to our expectations, Opt67 (1st to 3rd) turned out to be infeasible (more precisely, 
it had a negative minimum). This was caused by the lower capacity constraints~(\ref{lower-bound}).
If we remove the lower capacity constraints, 
we can assign all students to classes up to their third choices. 
As a necessary condition for the existence of a feasible solution, it must satisfy
\begin{equation}
\sum_{i = 1}^3 n_{si} \geq 7,
\label{necessary}
\end{equation}
for all $s \in {\mathcal{S}}$, where $n_{si}$ represents the number of students whose $i$-th choices are $s$. 
Nevertheless, two classes break the condition~(\ref{necessary}). 
That is, the reason why some students are assigned to their fourth choice is 
the policy of the university, which wants to prevent the occurrence of depopulated classes.
\par
In order to reduce the number of students who receive inopportune results,
we applied well-known approaches in the study of profile-based matchings (see Table~\ref{Tbl:additional_R1}). 
The first is the rank-maximal matching approach~\cite{huang2019exact, irving2006rank, kavitha2006efficient, michail2007reducing, paluch2013capacitated}. A maximum-cardinality matching is said to be rank-maximal 
if it lexicographically maximizes the number of students 
assigned to their first choices, then  
second choices, and so on. We can obtain a rank-maximal matching by solving $\mathrm{CA}(\bs{p})$ for 
$\bs{p} = (N^3, N^2, N, 1, 0, -L)$, where $N$ and $L$ are sufficiently large integers and we set 
\begin{align*}
N &:= |\mathcal{S}| + 1, \\
L &:= N^3|\mathcal{S}|+1.
\end{align*}
The second is the fair matching approach~\cite{huang2016fair}. A (maximum-cardinality) matching is said to be fair 
if it lexicographically minimizes the number of students 
assigned to classes not included in their choices, the number of students assigned to their 
fifth choices, and so on. 
We can obtain a fair matching by solving $\mathrm{CA}(\bs{p})$ for 
$\bs{p} = (0, -1, -N, -N^2, -N^3, -N^4)$. 
\par 
The third approach is our own method. 
We propose a compromise between maximum-utility matchings and fair matchings. 
Specifically, we lexicographically minimize 
the number of students 
assigned to classes not included in their choices, the number of students assigned to their 
fifth choices, and the number of students assigned to their 
fourth choices. In addition, we maximize the sum of the utilities of students 
assigned to their first to third choices with respect to the utility vector of Opt67. 
We can realize this by solving $\mathrm{CA}(\bs{p})$ for 
$\bs{p} = (100, 67, 45, -M, -MN, -MN^2)$. 
We denote this model as Opt67$\times$Fair.

\begin{table}[H]
\centering
\caption{Three profile-based models}
\label{Tbl:additional_R1}
\scalebox{1.0}{
\begin{tabular}{|l|l|} \hline
Model &  Utility vector $\bs{p}$ \\ \hline
Rank-maximal & $(N^3, N^2, N, 1, 0, -L)$ \\
Fair & $(0, -1, -N, -N^2, -N^3, -N^4)$ \\
Opt67$\times$Fair & $(100, 67, 45, -M, -MN, -MN^2)$ \\ \hline
\end{tabular}
}
\end{table}

\begin{figure}[H]
\centering
\includegraphics[width = 120mm]{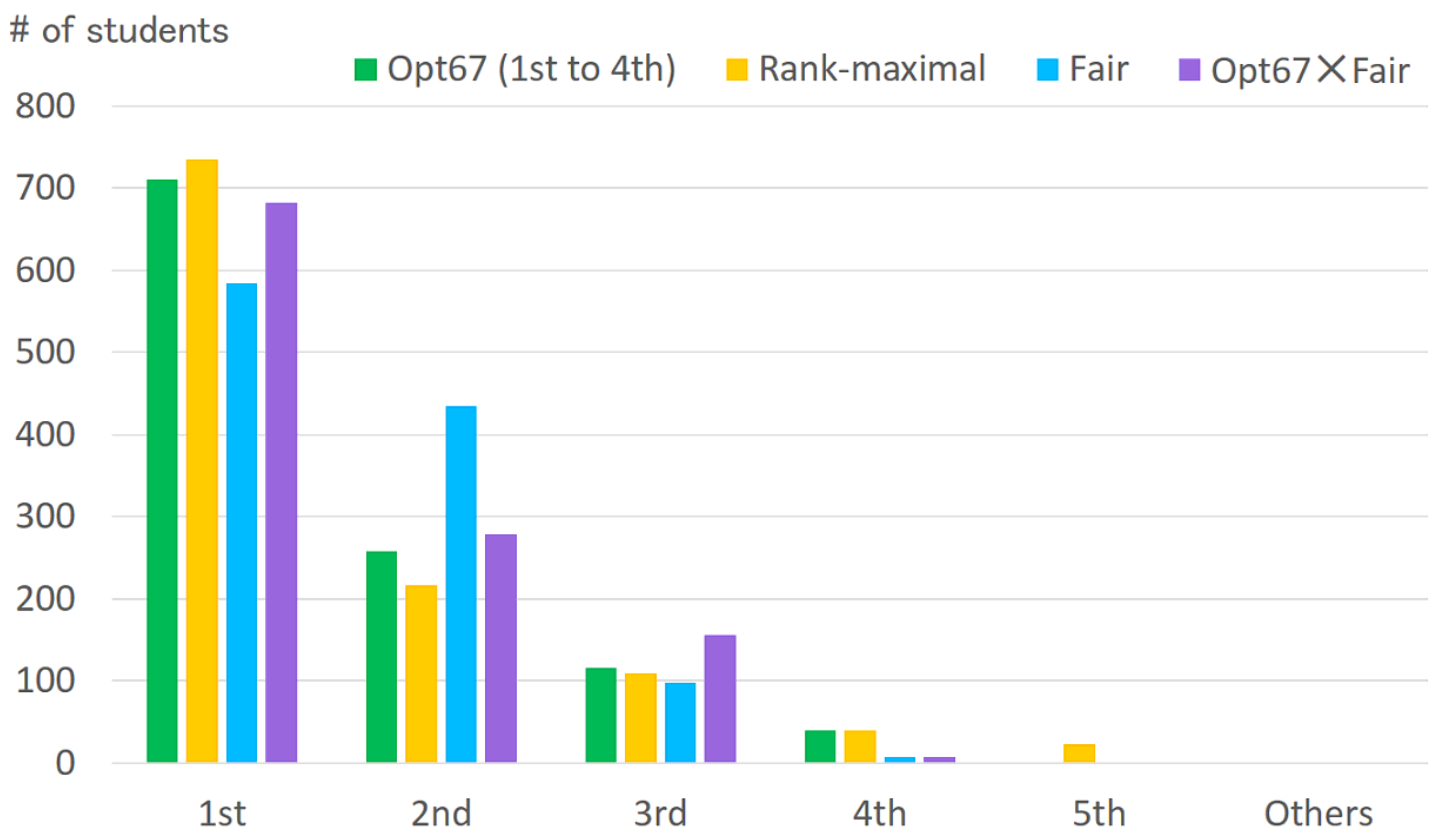}
\caption{Matching results by the profile-based models in FY2019}
\label{Fig:results2_R1}
\end{figure}

The results are shown in Table~\ref{Tbl:results2_R1} and Figure~\ref{Fig:results2_R1}. 
The final assignment is shown in Table~\ref{Tbl:final_R1}.

\begin{table}[H]
\centering
\caption{Matching results by the profile-based models in FY2019}
\label{Tbl:results2_R1}
\scalebox{1.0}{
\begin{tabular}{|c|c|c|c|c|c|c|c|c|} \hline
\multirow{2}{*}{Model}  &  \multicolumn{6}{c|}{\# of students (Total = 1123)}  &  \multicolumn{2}{c|}{Average}  \\ \cline{2-9}
  &  1st  &  2nd  &  3rd  &  4th  &  5th  &  Others  &  Utility (Opt67) &  Rank \\ \hline
Rank-maximal  &  736  &  216  & 109  &  39  &  23  &  0  &  84.24   &  1.572  \\  
Fair  &  584  &  435  &  97  &  7  &  0  &  0  &  82.03   &  1.579  \\
Opt67$\times$Fair   &  682  &  278  &  156  &  7  &  0  &  0  &  83.75   &  1.544 \\ \hline
\end{tabular}
}
\end{table}

\begin{table}[p]
\centering
\caption{Final assignment by Opt67$\times$Fair in FY2019}
\label{Tbl:final_R1}
\scalebox{0.85}{
\begin{tabular}{|r|r|r|r|r|r|r|r|r|r|r|r|r|r|} \hline
\multirow{2}{*}{Class} & \multicolumn{2}{c|}{Capacity}  &  \multicolumn{5}{c|}{\cg \# of students assigned} &  \multicolumn{6}{c|}{\cg \# of students ranking the class} \\ \cline{2-3} \cline{5-8} \cline{10-14}
 & Lower & Upper & \cg Total & E & M & SS & ST & \cg Total & 1st & 2nd & 3rd & 4th & 5th \\ \hline 
1 & 7 & 25 & \cg 25 & 15 & 2 & 4 & 4 & \cg 143 & 15 & 29 & 27 & 41 & 31 \\ 
2 & 7 & 25 & \cg 18 & 0 & 2 & 4 & 12 & \cg 58 & 8 & 6 & 11 & 19 & 14 \\ 
3 & 7 & 30 & \cg 11 & 2 & 2 & 1 & 6 & \cg 31 & 1 & 8 & 8 & 8 & 6 \\ 
4 & 7 & 25 & \cg 25 & 4 & 6 & 2 & 13 & \cg 104 & 14 & 19 & 12 & 21 & 38 \\ 
5 & 7 & 25 & \cg 13 & 0 & 1 & 0 & 12 & \cg 54 & 2 & 6 & 16 & 11 & 19 \\ 
6 & 7 & 36 & \cg 36 & 20 & 2 & 0 & 14 & \cg 90 & 18 & 24 & 18 & 18 & 12 \\ 
7 & 7 & 40 & \cg 40 & 2 & 18 & 1 & 19 & \cg 123 & 9 & 23 & 31 & 31 & 29 \\ 
8 & 7 & 32 & \cg 18 & 2 & 0 & 0 & 16 & \cg 43 & 5 & 6 & 9 & 9 & 14 \\ 
9 & 7 & 25 & \cg 16 & 0 & 0 & 14 & 2 & \cg 44 & 10 & 8 & 9 & 9 & 8 \\ 
10 & 7 & 25 & \cg 25 & 1 & 2 & 0 & 22 & \cg 70 & 13 & 11 & 17 & 18 & 11 \\ 
11 & 7 & 20 & \cg 13 & 0 & 1 & 2 & 10 & \cg 70 & 6 & 11 & 11 & 12 & 30 \\ 
12 & 7 & 40 & \cg 40 & 0 & 14 & 1 & 25 & \cg 203 & 28 & 51 & 43 & 46 & 35 \\ 
13 & 7 & 40 & \cg 40 & 2 & 29 & 0 & 9 & \cg 241 & 63 & 65 & 38 & 42 & 33 \\ 
14 & 7 & 25 & \cg 25 & 0 & 10 & 0 & 15 & \cg 339 & 71 & 68 & 96 & 57 & 47 \\ 
15 & 7 & 25 & \cg 25 & 3 & 4 & 0 & 18 & \cg 155 & 37 & 35 & 29 & 23 & 31 \\ 
16 & 7 & 30 & \cg 18 & 2 & 0 & 5 & 11 & \cg 55 & 8 & 15 & 9 & 17 & 6 \\ 
17 & 7 & 16 & \cg 16 & 0 & 1 & 0 & 15 & \cg 70 & 16 & 12 & 11 & 14 & 17 \\ 
18 & 7 & 40 & \cg 33 & 6 & 4 & 2 & 21 & \cg 104 & 11 & 19 & 21 & 16 & 37 \\ 
19 & 7 & 18 & \cg 18 & 11 & 1 & 2 & 4 & \cg 45 & 9 & 8 & 12 & 8 & 8 \\ 
20 & 7 & 25 & \cg 25 & 2 & 2 & 7 & 14 & \cg 82 & 15 & 16 & 14 & 25 & 12 \\ 
21 & 7 & 25 & \cg 17 & 12 & 4 & 1 & 0 & \cg 67 & 7 & 8 & 20 & 20 & 12 \\ 
22 & 7 & 25 & \cg 25 & 1 & 2 & 1 & 21 & \cg 190 & 61 & 38 & 23 & 19 & 49 \\ 
23 & 7 & 25 & \cg 24 & 10 & 2 & 9 & 3 & \cg 56 & 12 & 9 & 16 & 9 & 10 \\ 
24 & 7 & 12 & \cg 7 & 1 & 3 & 0 & 3 & \cg 13 & 0 & 4 & 4 & 4 & 1 \\ 
25 & 7 & 25 & \cg 25 & 1 & 21 & 0 & 3 & \cg 171 & 46 & 43 & 29 & 32 & 21 \\ 
26 & 7 & 25 & \cg 7 & 3 & 0 & 2 & 2 & \cg 48 & 3 & 9 & 6 & 18 & 12 \\ 
27 & 7 & 20 & \cg 20 & 9 & 1 & 3 & 7 & \cg 116 & 16 & 21 & 29 & 22 & 28 \\ 
28 & 7 & 30 & \cg 30 & 10 & 7 & 0 & 13 & \cg 140 & 29 & 46 & 23 & 21 & 21 \\ 
29 & 7 & 25 & \cg 8 & 3 & 0 & 0 & 5 & \cg 45 & 2 & 6 & 9 & 11 & 17 \\ 
30 & 7 & 40 & \cg 40 & 12 & 17 & 2 & 9 & \cg 252 & 97 & 40 & 36 & 44 & 35 \\ 
31 & 7 & 20 & \cg 20 & 4 & 12 & 0 & 4 & \cg 164 & 41 & 29 & 31 & 28 & 35 \\ 
32 & 7 & 16 & \cg 9 & 1 & 0 & 0 & 8 & \cg 39 & 2 & 8 & 9 & 8 & 12 \\ 
33 & 7 & 25 & \cg 25 & 9 & 7 & 3 & 6 & \cg 84 & 9 & 12 & 17 & 15 & 31 \\ 
34 & 7 & 25 & \cg 23 & 8 & 7 & 1 & 7 & \cg 78 & 5 & 19 & 17 & 17 & 20 \\ 
35 & 7 & 25 & \cg 25 & 6 & 9 & 1 & 9 & \cg 199 & 17 & 25 & 49 & 65 & 43 \\ 
36 & 7 & 15 & \cg 15 & 3 & 2 & 1 & 9 & \cg 98 & 18 & 19 & 21 & 21 & 19 \\ 
37 & 7 & 20 & \cg 7 & 1 & 2 & 3 & 1 & \cg 25 & 1 & 3 & 1 & 9 & 11 \\ 
38 & 7 & 25 & \cg 25 & 1 & 12 & 0 & 12 & \cg 375 & 131 & 92 & 52 & 50 & 50 \\ 
39 & 7 & 30 & \cg 13 & 1 & 0 & 1 & 11 & \cg 61 & 9 & 9 & 10 & 15 & 18 \\ 
40 & 7 & 25 & \cg 25 & 1 & 2 & 5 & 17 & \cg 72 & 21 & 10 & 21 & 10 & 10 \\ 
41 & 7 & 30 & \cg 30 & 18 & 2 & 2 & 8 & \cg 87 & 16 & 26 & 17 & 13 & 15 \\ 
42 & 7 & 25 & \cg 25 & 13 & 0 & 2 & 10 & \cg 112 & 21 & 25 & 20 & 28 & 18 \\ 
43 & 7 & 20 & \cg 20 & 2 & 2 & 0 & 16 & \cg 206 & 19 & 37 & 50 & 40 & 60 \\ 
44 & 7 & 20 & \cg 7 & 0 & 3 & 1 & 3 & \cg 9 & 1 & 0 & 1 & 5 & 2 \\ 
45 & 7 & 35 & \cg 35 & 1 & 22 & 2 & 10 & \cg 185 & 13 & 34 & 49 & 44 & 45 \\ 
46 & 7 & 40 & \cg 40 & 0 & 8 & 1 & 31 & \cg 214 & 77 & 48 & 46 & 26 & 17 \\ 
47 & 7 & 30 & \cg 11 & 0 & 0 & 7 & 4 & \cg 41 & 7 & 3 & 13 & 6 & 12 \\ 
48 & 7 & 20 & \cg 20 & 4 & 3 & 4 & 9 & \cg 83 & 7 & 13 & 15 & 27 & 21 \\ 
49 & 7 & 40 & \cg 40 & 10 & 14 & 3 & 13 & \cg 141 & 32 & 28 & 30 & 34 & 17 \\ 
50 & 7 & 25 & \cg 25 & 12 & 2 & 1 & 10 & \cg 120 & 44 & 19 & 17 & 17 & 23 \\ 
\hline
\end{tabular}
} \\
{\footnotesize 
\begin{itemize}
\item E = Faculty of Education, M = Faculty of Medicine, SS = Faculty of Social and Information Studies, 
ST = School of Science and Technology
\item The order of the classes has been shuffled to ensure anonymity.
\end{itemize}
}
\end{table}

Finally, we could reduce the number of students assigned to their fourth choices to seven. 
After discussing these results with the members concerned, 
they agreed that either the Fair category or Opt67$\times$Fair can be adopted; however  
it is difficult to determine which of the two is preferable.
Thus, Opt67$\times$Fair was adopted by a majority vote among the members.
%
\section{Concluding Remarks}
%
Through a two-year case study of the class assignment problem at Gunma University, 
we have augmented the optimization approach introduced by Konno~\cite{konno1992introduction} 
and applied them to real field data. 
Our own versions include the minimax-rank constrained maximum-utility matching and a 
compromise between maximum-utility matchings and fair matchings.
Our results show that we have succeeded in reducing the percentage of students not assigned to classes 
up to their third choices from twenty percent to almost zero (zero percent in FY2018 and 0.62 percent in FY2019).
In addition, we also could observe the potential inefficiency of the student proposing deferred acceptance 
mechanism with single tie-breaking, which is well known in the literature on the school choice problem. 
We believe that there are no perfect model for the class assignment problem, 
and there is still scope for discussion and further adjustments depending on real field data provided.
%
\section*{Acknowledgments}
%
The authors would like to thank Dr. Kenji Kubota, Executive Vice President of Gunma University, 
who asked us to improve the class assignment procedure.
The authors are also deeply grateful to all the staffs of the Liberal Arts Education Section, 
Gunma University. 
Without their cooperation, the optimization techniques would never have been used in 
university, and 
this investigation would not have been possible. 
We would like to thank Editage (www.editage.com) for English language editing.
\par
Akifumi Kira was supported in part by JSPS KAKENHI Grant Numbers 17K12644, Japan.
Kiyohito Nagano was supported by JSPS KAKENHI Grant Numbers 17K00036, Japan.
Manabu Sugiyama was supported by JSPS KAKENHI Grant Numbers 20K01847, Japan.
Naoyuki Kamiyama was supported in part by JST, PRESTO Grant Number JPMJPR1753, Japan. 

\begin{thebibliography}{99}
\bibitem{abdulkadiroglu2011resolving}
A.~Abdulkadiro\u{g}lu,  Y.-K.~Che, and Y.~Yasuda,
Resolving conflicting preferences in school choice: The \lq\lq boston mechanism" reconsidered, 
{\it American Economic Review}, {\bf 101}(1), 2011, 399--410.

\bibitem{abdulkadiroglu2009strategy}
A.~Abdulkadiro\u{g}lu,  P.~Pathak, and A.E~Roth,
Strategy-proofness versus efficiency in matching with indifferences: Redesigning the NYC High School Match, 
{\it American Economic Review}, {\bf 99}(5), 2009, 1954--1978.

\bibitem{abdulkadiroglu2003school}
A.~Abdulkadiro\u{g}lu and T.~S{\"o}nmez,
School choice: A mechanism design approach, 
{\it American Economic Review}, {\bf 93}(3), 2003, 729--747.

\bibitem{biro2010college}
P.~Bir\`{o}, T.~Fleiner, R.W.~Irving, and D.~Manlove, 
The college admissions problem with lower and common quotas. 
{\it Theoret. Comput. Sci.}
{\bf 411}(34--36), 2010, 3136--3153.

\bibitem{conforti2014integer}
M.~Conforti, G.~Cornu{\'e}jols, and G.~Zambelli, 
\newblock {\em Integer Programming}
\newblock (Springer, 2014).

\bibitem{erdil2008what}
A.~Erdil, H.~Ergin, 
What's the matter with tie-breaking? Improving efficiency in school choice, 
{\it American Economic Review}, {\bf 98}(3), 2008, 669--689.

\bibitem{ergin2006games}
H.~Ergin and T.~S{\"o}nmez, 
Games of school choice under the Boston mechanism, 
{\it Journal of public Economics}, {\bf 90}(1-2), 2006, 215--237.

\bibitem{fleiner2016matroid}
T.~Fleiner and N.~Kamiyama, 
A matroid approach to stable matchings with lower quotas,
{\it Mathematics of Operations Research}, 
{\bf 41}(2), 2016, 734--744.

\bibitem{fragiadakis2016}
D.~Fragiadakis, A.~Iwasaki, P.~Troyan, S.~Ueda, and M.~Yokoo, Makoto, 
Strategyproof matching with minimum quotas, 
{\it ACM Trans. Econ. Comput.}, 
{\bf 4}(1), 2016, 2167--8375.

\bibitem{gale1962college}
D.~Gale and L.S.~Shapley, 
College admissions and the stability of marriage, 
{\it The American Mathematical Monthly}, {\bf 69}(1), 1962, 9--15.

\bibitem{hagberg2008exploring}
A.A.~Hagberg, D.A.~Schult and P.J.~Swart, 
Exploring network structure, dynamics, and function using NetworkX, 
In: G.~Varoquaux, T.~Vaught, and J.~Millman (Eds), 
{\em Proceedings of the 7th Python in Science Conference (SciPy2008)}, 
(Pasadena, CA USA), 11--15, Aug 2008.

\bibitem{hamada2016hospitals}
K.~Hamada, K.~Iwama, S.~Miyazaki, 
The hospitals/residents problem with lower quotas,  
{\it Algorithmica} {\bf 74}(1): 2016, 440--465.

\bibitem{huang2010classified}
C.-C.~Huang,  
Classified stable matching,
{\em Proc. 21st Annual ACM-SIAM Sympos. on Discrete Algorithms (SODA'10)}, 
2010, 1235--1253.

\bibitem{huang2019exact}
C.-C. Huang, N.~Kakimura, and N.~Kamiyama, 
Exact and approximation algorithms for weighted matroid intersection, 
{\it Mathematical Programming}, {\bf 177}(1--2), 2019, 85--112.

\bibitem{huang2016fair}
C.-C.~Huang, T.~Kavitha, K.~Mehlhorn, and D.~Michail, 
Fair matchings and related problems, 
{\it Algorithmica}, {\bf 74}(3), 2016, 1184--1203.

\bibitem{irving2006rank}
R.W.~Irving, T.~Kavitha, K.~Mehlhorn, D.~Michail, and K.E.~Paluch, 
Rank-maximal matchings, 
{\it ACM Trans. Algorithms} {\bf 2}(4), 2006, 602--610.

\bibitem{kavitha2006efficient}
T.~Kavitha and C.D.~Shah, 
Effcient algorithms for weighted rank-maximal matchings and related problems, 
In: {\it Proceedings of the 17th ISAAC}, 2006, 153--162.

\bibitem{kira2017indirect}
A.~Kira, H.~Iwane, H.~Anai, Y.~Kimura, and K.~Fujisawa, 
An indirect search algorithm for disaster restoration with precedence and synchronization constraints, 
{\it Pacific Journal of Mathematics for Industry}, {\bf 9:}7, 2017, 15 pages.

\bibitem{kira2018nursery}
A.~Kira, N.~Kamiyama, H.~Iwashita, K.~Ohori, and H.~Anai, 
Nursery school matching with siblings -- striving for fairness by game theory, 
{\it J-LIS: Local Government Information Magazine}, {\bf 4}(10), 2018, 32--37. (in Japanese)

\bibitem{konno1992introduction}
H.~Konno,
\newblock {\em Introduction to Mathematical Decision Making: Campus OR}
\newblock (Asakura Publishing, 1992). (in Japanese)

\bibitem{konno1991optimal}
H.~Konno and Z.~Zhu,
The optimal class assignment problem: a case study at Tokyo Institute of Technology, 
{\it Communications of Operations Research of Japan}, {\bf 36}, 
1991, 85--89. (in Japanese)

\bibitem{michail2007reducing}
D.~Michail,  
Reducing rank-maximal to maximum weight matching, 
{\it Theor. Comput. Sci.} {\bf 389}(1--2), 2007, 125--132. 

\bibitem{ohori2018mathematical}
K.~Ohori and H.~Anai, 
Mathematical technologies and artificial intelligence toward human-centric innovation, 
In: {\it Innovative Approaches in Agent-Based Modelling and Business Intelligence}, 
2018, Springer, 9--22.

\bibitem{paluch2013capacitated}
K.~Paluch, 
Capacitated rank-maximal matchings, 
In {\it Proceedings of the 8th CIAC}, 2013, 324--335. 

\bibitem{schrijver1998theory}
A.~Schrijver,
\newblock {\em Theory of Linear and Integer Programming}
\newblock (John Wiley \& Sons, 1998).

\bibitem{yamada2017modeling}
H.~Yamada, K.~Ohori, T.~Iwao, A.~Kira, N.~Kamiyama, H.~Yoshida, and H.~Anai, 
Modeling and managing airport passenger flow under uncertainty: 
a case of Fukuoka Airport in Japan, 
In: {\it Social Informatics (Proceedings of SocInfo 2017), Lecture Note in Computer Science (LNCS)}, 
{\bf 10540}, 2017, Springer, 419--430.

\bibitem{yokoi2017generalized}
Y.~Yokoi, 
A generalized polymatroid approach to stable matchings with lower quotas, 
{\it Mathematics of Operations Research}, 
{\bf 42}(1), 2017, 238--255.
\end{thebibliography}


\end{document}